\documentclass[12pt,twoside,english]{article}
\usepackage[T1]{fontenc}
\usepackage[latin1]{inputenc}
\usepackage{geometry}
\usepackage{amssymb}

\makeatletter

\usepackage{babel,cite}
\def\p{\partial}

\usepackage{babel}
\makeatother
\begin{document}

\renewcommand{\title}[1]{\vspace{10mm}\noindent{\Large{\bf #1}}\vspace{8mm}} \newcommand{\authors}[1]{\noindent{\large #1}\vspace{5mm}} \newcommand{\address}[1]{{\itshape #1\vspace{2mm}}} 

\begin{titlepage}

\begin{flushright}
LMU-ASC 69/05\\
MPP-2005-120
\end{flushright}

\begin{center}    \title{ \LARGE Noncommutative Spaces and Gravity \vspace{1cm} }

\authors{Frank {\sc Meyer} \vspace{0.8cm}}

\address{Max--Planck--Institut f\"ur Physik\\ \it (Werner-Heisenberg Institut) \\ \it F\"ohringer Ring 6, D-80805 M\"unchen, Germany \\}

\address{Arnold Sommerfeld Center, Department f\"ur Physik\\ \it Ludwig-Maximilians-Universit\"at M\"unchen, Theresienstra{\ss}e 37\\ \it D-80333 M\"unchen, Germany \vspace{0,5cm}}

\address{{\it E--mail:} {\tt meyerf@theorie.physik.uni-muenchen.de}}

\vskip 0,4cm

\textbf{Abstract}

\vskip 3mm

\begin{minipage}{14cm}%

We give an introduction to an algebraic construction of a gravity theory on noncommutative spaces which is based on a deformed algebra of (infinitesimal) diffeomorphisms. We start with some fundamental ideas and concepts of noncommutative spaces. Then the $\theta$-deformation of diffeomorphisms is studied and a tensor calculus is defined. A deformed Einstein-Hilbert action invariant with respect to deformed diffeomorphisms is given. Finally, all noncommutative fields are expressed in terms of their commutative counterparts up to second order of the deformation parameter  using the $\star$-product. This allows to study explicitly deviations to Einstein's gravity theory in orders of $\theta$. This lecture is based on joined work with P. Aschieri, C. Blohmann, M. Dimitrijevi{\' c}, P. Schupp and J. Wess.

\end{minipage}

\vspace{0,1cm}

\end{center}

\address{Based on talks given at the First Modave Summer School in Mathematical Physics,}

\vspace{-2mm}

\address{June 2005, Modave (Belgium); HEP 2005, July 2005, Lisboa (Portugal);}

\vspace{-2mm}

\address{XXVIII Spanish Relativity Meeting, September 2005, Oviedo (Spain)}

\end{titlepage}

\newpage

\tableofcontents{}

\section{Noncommutative Spaces\label{sec:Noncommutative-Spaces}}

In field theories one usually considers differential space-time manifolds.
In the noncommutative realm, the notion of a point is no longer well-defined
and we have to give up the concept of differentiable manifolds. However,
the space of functions on a manifold is an algebra. A generalization
of this algebra can be considered in the noncommutative case. We take
the algebra freely generated by the \emph{noncommutative coordinates}
$\hat{x}^{i}$ which respects commutation relations of the type 

\begin{equation}
{}[\hat{x}^{\mu},\hat{x}^{\nu}]=C^{\mu\nu}(\hat{x})\neq0\,.\label{eq: comm rel}\end{equation}
 Without bothering about convergence, we take the space of formal
power series in the coordinates $\hat{x}^{i}$ and divide by the ideal
generated by the above relations\[
\widehat{\mathcal{A}}_{\hat{x}}=\mathbb{C}\langle\langle\hat{x}^{0},\dots,\hat{x}^{n}\rangle\rangle/([\hat{x}^{\mu},\hat{x}^{\nu}]-C^{\mu\nu}(\hat{x}))\,.\]
The function $C^{\mu\nu}(\hat{x})$ is unknown. For physical reasons
it should be a function that vanishes at large distances where we
experience the commutative world and may be determined by experiments.
Nevertheless, one can consider a power-series expansion \[
C^{\mu\nu}(\hat{x})=i\theta^{\mu\nu}+iC^{\mu\nu}{}_{\rho}\hat{x}^{\rho}+(q\hat{R}^{\mu\nu}{}_{\rho\sigma}-\delta_{\rho}^{\nu}\delta_{\sigma}^{\mu})\hat{x}^{\rho}\hat{x}^{\sigma}+\dots,\]
 where $\theta^{\mu\nu}$, $C^{\mu\nu}{}_{\rho}$ and $q\hat{R}^{\mu\nu}{}_{\rho\sigma}$
are constants, and study cases where the commutation relations are
constant, linear or quadratic in the coordinates. At very short distances
those cases provide a reasonable approximation for $C^{\mu\nu}(\hat{x})$
and lead to the following three structures

\begin{enumerate}
\item canonical or $\theta$-deformed case: \begin{equation}
[\hat{x}^{\mu},\hat{x}^{\nu}]=i\theta^{\mu\nu}.\label{eq: canoncial strucutre}\end{equation}
 
\item Lie algebra case: \begin{equation}
[\hat{x}^{\mu},\hat{x}^{\nu}]=iC^{\mu\nu}{}_{\rho}\hat{x}^{\rho}.\label{eq: Lie structure}\end{equation}
 
\item Quantum Spaces: \begin{equation}
\hat{x}^{\mu}\hat{x}^{\nu}=q\hat{R}^{\mu\nu}{}_{\rho\sigma}\hat{x}^{\rho}\hat{x}^{\sigma}.\label{eq: quantum structure}\end{equation}
 
\end{enumerate}
We denote the algebra generated by noncommutative coordinates $\hat{x}^{\mu}$
which are subject to the relations (\ref{eq: canoncial strucutre})
by $\hat{\mathcal{A}}$. We shall often call it the \emph{algebra
of noncommutative functions}. Commutative functions will be denoted
by $\mathcal{A}$. In what follows we will exclusively consider the
$\theta$-deformed case (\ref{eq: canoncial strucutre}) but we note
that the algebraic construction presented here can be generalized
to more complicated noncommutative structures of the above type which
possess the Poincar{\' e}-Birkhoff-Witt (PBW) property. The PBW-property
states that the space of polynomials in noncommutative coordinates
of a given degree is isomorphic to the space of polynomials in the
commutative coordinates. Such an isomorphism between polynomials of
a fixed degree is given by an ordering prescription. One example is
the \emph{symmetric ordering} (or Weyl-ordering) $W$ which assigns
to any monomial the totally symmetric ordered monomial \begin{eqnarray}
W:\,\mathcal{A} & \rightarrow & \hat{\mathcal{A}}\nonumber \\
x^{\mu} & \mapsto & \hat{x}^{\mu}\label{eq: Weyl ordering}\\
x^{\mu}x^{\nu} & \mapsto & \frac{1}{2}(\hat{x}^{\mu}\hat{x}^{\nu}+\hat{x}^{\nu}\hat{x}^{\mu})\nonumber \\
 & \cdots & .\nonumber \end{eqnarray}

To study the dynamics of fields we need a differential calculus on
the noncommutative algebra $\hat{\mathcal{A}}$. Derivatives are maps
on the deformed coordinate space \cite{Wess:2003da}\[
\hat{\partial}_{\mu}:\,\hat{\mathcal{A}}\rightarrow\hat{\mathcal{A}}\,.\]
This means that they have to be consistent with the commutation relations
of the coordinates. In the $\theta$-constant case a consistent differential
calculus can be defined very easily by\emph{}%
\footnote{We use brackets to distinguish the action of a differential operator
from the multiplication in the algebra of differential operators.%
} \begin{eqnarray}
[\hat{\partial}_{\mu},\hat{x}^{\nu}]=\delta_{\mu}^{\nu} &  & (\hat{\partial}_{\mu}\hat{x}^{\nu})=\delta_{\mu}^{\nu}\nonumber \\
{}[\hat{\partial}_{\mu},\hat{\partial}_{\nu}]=0.\label{def: differential calculus}\end{eqnarray}
 It is the fully undeformed differential calculus. The above definitions
yield the usual Leibniz-rule for the derivatives $\hat{\partial}_{\mu}$\begin{equation}
(\hat{\partial}_{\mu}\hat{f}\hat{g})=(\hat{\partial}_{\mu}\hat{f})\hat{g}+\hat{f}(\hat{\partial}_{\mu}\hat{g}).\label{eq; Leibniz}\end{equation}
This is a special feature of the fact that $\theta^{\mu\nu}$ are
constants. In the more complicated examples of noncommutative structures
this undeformed Leibniz-rule usually cannot be preserved but one has
to consider deformed Leibniz-rules for the derivatives \cite{Wess:1991vh}.
Note that (\ref{def: differential calculus}) also implies that \begin{equation}
(\hat{\partial}_{\mu}\hat{f})=\widehat{(\partial_{\mu}f)}.\label{eq: hat d equals d hat}\end{equation}

The Weyl ordering (\ref{eq: Weyl ordering}) can be formally implemented
by the map \[
f\mapsto W(f)=\frac{1}{(2\pi)^{\frac{n}{2}}}\int d^{n}k\, e^{ik_{\mu}\hat{x}^{\mu}}\tilde{f}(k)\]
 where $\tilde{f}$ is the Fourier transform of $f$\[
\tilde{f}(k)=\frac{1}{(2\pi)^{\frac{n}{2}}}\int d^{n}x\, e^{-ik_{\mu}x^{\mu}}f(x).\]
 This is due to the fact that the exponential is a fully symmetric
function. Using the Baker-Campbell-Hausdorff formula one finds \begin{equation}
e^{ik_{\mu}\hat{x}^{\mu}}e^{ip_{\nu}\hat{x}^{\nu}}=e^{i(k_{\mu}+p_{\mu})\hat{x}^{\mu}-\frac{i}{2}k_{\mu}\theta^{\mu\nu}p_{\nu}}.\label{eq: BKH}\end{equation}
 This immediately leads to the following observation \begin{eqnarray}
\hat{f}\hat{g} & = & W(f)W(g)=\frac{1}{(2\pi)^{n}}\int d^{n}kd^{n}p\, e^{ik_{\mu}\hat{x}^{\mu}}e^{ip_{\nu}\hat{x}^{\nu}}\tilde{f}(k)\tilde{g}(p)\nonumber \\
 & = & \frac{1}{(2\pi)^{n}}\int d^{n}kd^{n}p\, e^{i(k_{\mu}+p_{\mu})\hat{x}^{\mu}}e^{-\frac{i}{2}k_{\mu}\theta^{\mu\nu}p_{\nu}}\tilde{f}(k)\tilde{g}(p)\nonumber \\
 & = & W(\mu\circ e^{\frac{i}{2}\theta^{\mu\nu}\partial_{\mu}\otimes\partial_{\nu}}f\otimes g),\label{eq: almost star product}\end{eqnarray}
where $\mu(f\otimes g):=fg$ is the multiplication map. 
 With (\ref{eq: hat d equals d hat}) we deduce from (\ref{eq: almost star product})
the equation \begin{equation}
\mu\circ e^{-\frac{i}{2}\theta^{\mu\nu}\hat{\partial}_{\mu}\otimes\hat{\partial}_{\nu}}\hat{f}\otimes\hat{g}=\widehat{fg}.\label{eq: comm pro expressed in NC alg}\end{equation}
 The above formula shows us how the commutative and the noncommutative
product are related. It will be important for us later on.

\section{Symmetries on Deformed Spaces\label{sec:Symmetries-on-Deformed}}

In general the commutation relations (\ref{eq: comm rel}) are not
covariant with respect to undeformed symmetries. For example the canonical
commutation relations (\ref{eq: canoncial strucutre}) break Lorentz
symmetry if we assume that the noncommutativity parameters $\theta^{\mu\nu}$
do not transform. 

The question arises whether we can \emph{deform} the symmetry in such
a way that it acts consistently on the deformed space (i.e. leaves
the deformed space invariant) and such that it reduces to the undeformed
symmetry in the commutative limit. The answer is yes: Lie algebras
can be deformed in the category of Hopf algebras (Hopf algebras coming
from a Lie algebra are also called Quantum Groups)%
\footnote{To be more precise the universal enveloping algebra of a Lie algebra can be deformed. The universal enveloping algebra of any Lie algebra is a Hopf algebra and this gives rise to deformations in the category of Hopf algebras.}. Important examples of such deformations are $q$-deformations: Drinfeld
and Jimbo have shown that there exists a $q$-deformation of the universal
enveloping algebra of an arbitrary semisimple Lie algebra\footnote{It is called $q$-deformation since it is a deformation in terms of a parameter $q$.}. Module
algebras of this $q$-deformed universal enveloping algebras are noncommutative
spaces with commutation relations of type (\ref{eq: quantum structure}).
There exists also a so-called $\kappa$-deformation of the Poincar{\' e}
algebra \cite{Dimitrijevic:2003wv,Dimitrijevic:2003pn} which leads
to a noncommutative space of the Lie type (\ref{eq: Lie structure}).
A Hopf algebra symmetry acting on the $\theta$-deformed space was
for a long time unknown. But recently also a $\theta$-deformation
of the Poincar{\' e} algebra leading to the algebra (\ref{eq: canoncial strucutre})
was constructed \cite{Oeckl:2000eg,Chaichian:2004za,Koch:2004ud,Aschieri:2005yw}. 

Quantum group symmetries lead to new features of field theories on
noncommutative spaces. Because of its simplicity, $\theta$-deformed
spaces are very well-suited to study those. First results on the consequences
of the $\theta$-deformed Poincar{\' e} algebra have already been
obtained \cite{Chaichian:2004za,Aschieri:2005yw}. However, it remains
unknown and subject of future investigations in which precise way
this recently discovered quantum group symmetry restricts the degrees
of freedom of the noncommutative field theory.

In the following we will construct explicitly a $\theta$-deformed
version of diffeomorphisms which consistently act on the noncommutative
space (\ref{eq: canoncial strucutre}). Then we present a gravity
theory which is invariant with respect to this deformed diffeomorphisms \cite{Aschieri:2005yw,Aschieri:2005zs}.

\section{Diffeomorphisms}

Gravity is a theory invariant with respect to diffeomorphisms. However,
to generalize the Einstein formalism to noncommutative spaces in order
to establish a gravity theory, it is important to first understand
that diffeomorphisms possess more mathematical structure than the
algebraic one: They are naturally equipped with a Hopf algebra structure.
In the common formulations of physical theories this additional Hopf
structure is hidden and does not play a crucial role. It is our aim
to deform the algebra of diffeomorphism in such a way that it acts
consistently on a noncommutative space. This can be done by exploiting
the full Hopf structure. In this section we first introduce the concept
of diffeomorphisms as Hopf algebra in the undeformed setting. 

Diffeomorphisms are generated by vector-fields $\xi$. Acting on functions,
vector-fields are represented as linear differential operators $\xi=\xi^{\mu}\partial_{\mu}$.
Vector-fields form a Lie algebra $\Xi$ over the field $\mathbb{C}$
with the Lie bracket given by \[
[\xi,\eta]=\xi\times\eta\]
 where $\xi\times\eta$ is defined by its action on functions \[
(\xi\times\eta)(f)=(\xi^{\mu}(\partial_{\mu}\eta^{\nu})\partial_{\nu}-\eta^{\mu}(\partial_{\mu}\xi^{\nu})\partial_{\nu})(f).\]
 The Lie algebra of \emph{infinitesimal diffeomorphisms} $\Xi$ can
be embedded into its universal enveloping algebra which we want to
denote by ${\cal U}(\Xi)$ . The universal enveloping algebra is
an associative algebra and possesses a natural Hopf algebra structure.
It is given by the following structure maps%
\footnote{The structure maps are defined on the generators $\xi\in\Xi$ and
the universal property of the universal enveloping algebra ${\cal U}(\Xi)$
assures that they can be uniquely extended as algebra homomorphisms
(respectively anti-algebra homomorphism in case of the antipode $S$)
to the whole algebra ${\cal U}(\Xi)$. %
}:

\begin{itemize}
\item An algebra homomorphism called \emph{coproduct} defined by \begin{eqnarray}
\Delta:\, {\cal U}(\Xi) & \rightarrow & {\cal U}(\Xi)\otimes {\cal U}(\Xi)\nonumber \\
\Xi\ni\xi & \mapsto & \Delta(\xi):=\xi\otimes1+1\otimes\xi.\label{def: coprod}\end{eqnarray}

\item An algebra homomorphism called \emph{counit} defined by \begin{eqnarray}
\varepsilon:\, {\cal U}(\Xi) & \rightarrow & \mathbb{C}\nonumber \\
\Xi\ni\xi & \mapsto & \varepsilon(\xi)=0.\label{def: epsilon}\end{eqnarray}

\item An anti-algebra homomorphism called \emph{antipode} defined by \begin{eqnarray}
S:\, {\cal U}(\Xi) & \rightarrow & {\cal U}(\Xi)\nonumber \\
\Xi\ni\xi & \mapsto & S(\xi)=-\xi.\label{def: antipod}\end{eqnarray}

\end{itemize}
For a precise definition and more details on Hopf algebras we refer
the reader to text books \cite{Chari-Presley,Klimyk:1997eb,Majid}.
For our purposes it shall be sufficient to note that the coproduct
implements how the Hopf algebra acts on a product in a representation
algebra (Leibniz-rule). Below we will make this more transparent.
It is now possible to study deformations of ${\cal U}(\Xi)$ in the category
of Hopf algebras. This leads to a deformed version of diffeomorphisms
- the fundamental building block of our approach to a gravity theory
on noncommutative spaces. Before studying this in detail, let us shortly
review the Einstein formalism. This way we first understand better
the meaning of the structure maps of a Hopf algebra introduced above.

Scalar fields are defined by their transformation property with respect
to infinitesimal coordinate transformations:\begin{equation}
\delta_{\xi}\phi=-\xi\phi=-\xi^{\mu}(\partial_{\mu}\phi).\label{def: scalar}\end{equation}
 The product of two scalar fields is transformed using the Leibniz-rule\begin{equation}
\delta_{\xi}(\phi\psi)=(\delta_{\xi}\phi)\psi+\phi(\delta_{\xi}\psi)=-\xi^{\mu}(\partial_{\mu}\phi\psi)\label{eq: Leibniz undeformed}\end{equation}
 such that the product of two scalar fields transforms again as a
scalar. The above Leibniz-rule can be understood in mathematical terms
as follows: The Hopf algebra ${\cal U}(\Xi)$ is represented on the space of
scalar fields by infinitesimal coordinate transformations $\delta_{\xi}$.
On scalar fields the action of $\delta_{\xi}$ is explicitly given
by the differential operator $-\xi^{\mu}\partial_{\mu}$. Of course,
the space of scalar fields is not only a vector space - it possesses
also an algebra structure - such as ${\cal U}(\Xi)$ is not only an algebra
but also a Hopf algebra - it possesses in addition the co-structure
maps defined above. We say that a Hopf algebra $H$ acts on an algebra
$A$ (or more precisely we say that $A$ is a left $H$-module algebra)
if $A$ is a module with respect to the algebra $H$ and if in addition
for all $h\in H$ and $a,b\in A$\begin{eqnarray}
h(ab) & = & \mu\circ\Delta h(a\otimes b)\label{def: H-module algebar}\\
h(1) & = & \varepsilon(h).\end{eqnarray}
Here $\mu$ is the multiplication map defined by $\mu(a\otimes b)=ab$.
In our concrete example where $H={\cal U}(\Xi)$ and $A$ is the algebra of
scalar fields we indeed have that the algebra of scalar fields is
a ${\cal U}(\Xi)-$module algebra. This can be seen easily if we rewrite (\ref{eq: Leibniz undeformed})
using (\ref{def: coprod}) for the generators $\xi\in\Xi$ for ${\cal U}(\Xi)$:\[
\delta_{\xi}(\phi\psi)=(\delta_{\xi}\phi)\psi+\phi(\delta_{\xi}\psi)=\mu\circ\Delta\xi(\phi\otimes\psi).\]
 It is also evident that \[
\delta_{\xi}1=0=\varepsilon(\xi)1.\]
 Now we are in the right mathematical framework: We study a Lie algebra
(here infinitesimal diffeomorphisms $\Xi$) and embed it in its universal
enveloping algebra (here ${\cal U}(\Xi)$). This universal enveloping algebra
is a Hopf algebra via a natural Hopf structure induced by (\ref{def: coprod},\ref{def: epsilon},\ref{def: antipod}). 

Physical quantities live in representations of this Hopf algebras.
For instance, the algebra of scalar fields is a ${\cal U}(\Xi)$-module algebra.
The action of ${\cal U}(\Xi)$ on scalar fields is given in terms of infinitesimal
coordinate transformations $\delta_{\xi}$. 

Similarly one studies tensor representations of ${\cal U}(\Xi)$. For example
vector fields are introduced by the transformation property \begin{eqnarray*}
\delta_{\xi}V_{\alpha} & = & -\xi^{\mu}(\partial_{\mu}V_{\alpha})-(\partial_{\alpha}\xi^{\mu})V_{\mu}\\
\delta_{\xi}V^{\alpha} & = & -\xi^{\mu}(\partial_{\mu}V^{\alpha})+(\partial_{\mu}\xi^{\alpha})V^{\mu}.\end{eqnarray*}
 The generalization to arbitrary tensor fields is straight forward:\begin{eqnarray*}
\delta_{\xi}T_{\nu_{1}\cdots\nu_{n}}^{\mu_{1}\cdots\mu_{n}} & = & -\xi^{\mu}(\partial_{\mu}T_{\nu_{1}\cdots\nu_{n}}^{\mu_{1}\cdots\mu_{n}})+(\partial_{\mu}\xi^{\mu_{1}})T_{\nu_{1}\cdots\nu_{n}}^{\mu\cdots\mu_{n}}+\cdots+(\partial_{\mu}\xi^{\mu_{n}})T_{\nu_{1}\cdots\nu_{n}}^{\mu_{1}\cdots\mu}\\
 &  & -(\partial_{\nu_{1}}\xi^{\nu})T_{\nu\cdots\nu_{n}}^{\mu_{1}\cdots\mu_{n}}-\cdots-(\partial_{\nu_{n}}\xi^{\nu})T_{\nu_{1}\cdots\nu}^{\mu_{1}\cdots\mu_{n}}.\end{eqnarray*}

As for scalar fields, we also find that the product of two tensors
transforms like a tensor. Summarizing, we have seen that scalar fields,
vector fields and tensor fields are representations of the Hopf algebra
${\cal U}(\Xi)$, the universal enveloping algebra of infinitesimal diffeomorphisms.
The Hopf algebra ${\cal U}(\Xi)$  acts via \emph{infinitesimal coordinate
transformations} $\delta_{\xi}$ which are subject to the relations:\begin{eqnarray}
[\delta_{\xi},\delta_{\eta}]=\delta_{\xi\times\eta} &  & \varepsilon(\delta_{\xi})=0\nonumber \\
\Delta\delta_{\xi}=\delta_{\xi}\otimes1+1\otimes\delta_{\xi} &  & S(\delta_{\xi})=-\delta_{\xi}.\label{eq: HA of diffeos}\end{eqnarray}
 The transformation operator $\delta_{\xi}$ is explicitly given by
differential operators which depend on the representation under consideration.
In case of scalar fields this differential operator is given by $-\xi^{\mu}\partial_{\mu}$.

\section{Deformed Diffeomorphisms}

The concepts introduced in the previous subsection can be deformed
in order to establish a consistent tensor calculus on the noncommutative
space-time algebra (\ref{eq: canoncial strucutre}). In this context
it is necessary to account the full Hopf algebra structure of the
universal enveloping algebra ${\cal U}(\Xi)$. 

In our setting the algebra $\hat{\mathcal{A}}$ possesses a noncommutative
product defined by \begin{equation}
[\hat{x}^{\mu},\hat{x}^{\nu}]=i\theta^{\mu\nu},\label{eq: NC space}\end{equation}
 We want to deform the structure maps (\ref{eq: HA of diffeos}) of
the Hopf algebra ${\cal U}(\Xi)$ in such a way that the resulting deformed
Hopf algebra which we denote by ${\cal U}(\hat{\Xi})$ consistently acts on
$\hat{\mathcal{A}}$. In the language introduced in the previous section
this means that we want $\hat{\mathcal{A}}$ to be a ${\cal U}(\hat{\Xi})$-module
algebra. We claim that the following deformation of ${\cal U}(\Xi)$ does the
job. Let ${\cal U}(\hat{\Xi})$ be generated as algebra by elements $\hat{\delta}_{\xi}$,
$\xi\in\Xi$. We leave the algebra relation undeformed and demand
\begin{equation}
[\hat{\delta}_{\xi},\hat{\delta}_{\eta}]=\hat{\delta}_{\xi\times\eta}\label{eq: Deform diffeos alg}\end{equation}
 but we deform the co-sector \begin{equation}
\Delta\hat{\delta}_{\xi}=e^{-\frac{i}{2}h\theta^{\rho\sigma}\hat{\partial}_{\rho}\otimes\hat{\partial}_{\sigma}}(\hat{\delta}_{\xi}\otimes1+1\otimes\hat{\delta}_{\xi})e^{\frac{i}{2}h\theta^{\rho\sigma}\hat{\partial}_{\rho}\otimes\hat{\partial}_{\sigma}},\label{eq: deform diffeos coprod}\end{equation}
 where \[
[\hat{\partial}_{\rho},\hat{\delta}_{\xi}]=\hat{\delta}_{(\partial_{\rho}\xi)}.\]
 The deformed coproduct (\ref{eq: deform diffeos coprod}) reduces
to the undeformed one (\ref{eq: HA of diffeos}) in the limit $\theta\rightarrow0$.
Antipode and counit remain undeformed \begin{eqnarray}
S(\hat{\delta}_{\xi})=-\hat{\delta}_{\xi} &  & \varepsilon(\hat{\delta}_{\xi})=0.\label{eq: deform diffeos antipode and counit}\end{eqnarray}
 We have to check whether the above deformation is a good one in the
sense that it leads to a consistent action on $\hat{\mathcal{A}}$.
First we need a differential operator acting on fields in $\hat{\mathcal{A}}$
which represents the algebra (\ref{eq: Deform diffeos alg}). Let
us consider the differential operator \begin{equation}
\hat{X}_{\xi}:=\sum_{n=0}^{\infty}\frac{1}{n!}(-\frac{i}{2})^{n}\theta^{\rho_{1}\sigma_{1}}\cdots\theta^{\rho_{n}\sigma_{n}}(\hat{\partial}_{\rho_{1}}\cdots\hat{\partial}_{\rho_{n}}\hat{\xi}^{\mu})\hat{\partial}_{\mu}\hat{\partial}_{\sigma_{1}}\cdots\hat{\partial}_{\sigma_{n}}.\label{def: X operator}\end{equation}
 This is to be understood like that: A vector-field $\xi=\xi^{\mu}\partial_{\mu}$
is determined by its coefficient functions $\xi^{\mu}$. In Section
\ref{sec:Noncommutative-Spaces} we saw that there is a vectorspace
isomorphism $W$ from the space of commutative to the space of noncommutative
functions which is given by the symmetric ordering prescription. The
image of a commutative function $f$ under the isomorphism $W$ is
denoted by $\hat{f}$ \[
W:\, f\mapsto W(f)=\hat{f}.\]
In (\ref{def: X operator}) $\hat{\xi}^{\mu}$ is therefore to be
interpreted as the image of $\xi^{\mu}$ with respect to $W$. Then
indeed we have \begin{equation}
[\hat{X}_{\xi},\hat{X}_{\eta}]=\hat{X}_{\xi\times\eta}.\label{eq: X repres vecfields}\end{equation}
 To see this we use result (\ref{eq: comm pro expressed in NC alg})
to rewrite $(\hat{X}_{\xi}\hat{\phi}):$\begin{eqnarray}
(\hat{X}_{\xi}\hat{\phi}) & = & \sum_{n=0}^{\infty}\frac{1}{n!}(-\frac{i}{2})^{n}\theta^{\rho_{1}\sigma_{1}}\cdots\theta^{\rho_{n}\sigma_{n}}(\hat{\partial}_{\rho_{1}}\cdots\hat{\partial}_{\rho_{n}}\hat{\xi}^{\mu})(\hat{\partial}_{\mu}\hat{\partial}_{\sigma_{1}}\cdots\hat{\partial}_{\sigma_{n}}\hat{\phi})\nonumber \\
 & = & \sum_{n=0}^{\infty}\frac{1}{n!}(-\frac{i}{2})^{n}\theta^{\rho_{1}\sigma_{1}}\cdots\theta^{\rho_{n}\sigma_{n}}(\hat{\partial}_{\rho_{1}}\cdots\hat{\partial}_{\rho_{n}}\hat{\xi}^{\mu})(\hat{\partial}_{\sigma_{1}}\cdots\hat{\partial}_{\sigma_{n}}\widehat{\partial_{\mu}\phi})\nonumber \\
 & = & \widehat{\xi^{\mu}(\partial_{\mu}\phi)}=\widehat{(\xi\phi)}.\label{eq: action undeformed}\end{eqnarray}
From (\ref{eq: action undeformed}) follows \[
(\hat{X}_{\xi}(\hat{X}_{\eta}\hat{\phi}))-(\hat{X}_{\eta}(\hat{X}_{\xi}\hat{\phi}))=\widehat{([\xi,\eta]\phi)}=(\hat{X}_{\xi\times\eta}\hat{\phi}),\]
which amounts to (\ref{eq: X repres vecfields}) and this is what
we wanted to show. 

It is therefore reasonable to introduce scalar fields $\hat{\phi}\in\hat{\mathcal{A}}$
by the transformation property \[
\hat{\delta}_{\xi}\hat{\phi}=-(\hat{X}_{\xi}\hat{\phi}).\]
 The next step is to work out the action of the differential operators
$\hat{X}_{\xi}$ on the product of two fields. A calculation \cite{Aschieri:2005yw}
shows that \[
(\hat{X}_{\xi}(\hat{\phi}\hat{\psi}))=\mu\circ(e^{-\frac{i}{2}h\theta^{\rho\sigma}\hat{\partial}_{\rho}\otimes\hat{\partial}_{\sigma}}(\hat{X}_{\xi}\otimes1+1\otimes\hat{X}_{\xi})e^{\frac{i}{2}h\theta^{\rho\sigma}\hat{\partial}_{\rho}\otimes\hat{\partial}_{\sigma}}\hat{\phi}\otimes\hat{\psi}).\]
 This means that the differential operators $\hat{X}_{\xi}$ act via
a \emph{deformed Leibniz rule} on the product of two fields. Comparing
with (\ref{eq: deform diffeos coprod}) we see that the deformed Leibniz
rule of the differential operator $\hat{X}_{\xi}$ is exactly the
one induced by the deformed coproduct (\ref{eq: deform diffeos coprod}):\[
\hat{\delta}_{\xi}(\hat{\phi}\hat{\psi})=e^{-\frac{i}{2}h\theta^{\rho\sigma}\hat{\partial}_{\rho}\otimes\hat{\partial}_{\sigma}}(\hat{\delta}_{\xi}\otimes1+1\otimes\hat{\delta}_{\xi})e^{\frac{i}{2}h\theta^{\rho\sigma}\hat{\partial}_{\rho}\otimes\hat{\partial}_{\sigma}}(\hat{\phi}\hat{\psi})=-\hat{X}_{\xi}\triangleright(\hat{\phi}\hat{\psi}).\]
 Hence, the deformed Hopf algebra ${\cal U}(\hat{\Xi})$ is indeed represented
on scalar fields $\hat{\phi}\in\hat{\mathcal{A}}$ by the differential
operator $\hat{X}_{\xi}$. The scalar fields form a ${\cal U}(\hat{\Xi})$-module
algebra. 

In analogy to the previous section we can introduce vector and tensor
fields as representations of the Hopf algebra ${\cal U}(\hat{\Xi})$. The transformation
property for an arbitrary tensor reads \begin{eqnarray*}
\hat{\delta}_{\xi}\hat{T}_{\nu_{1}\cdots\nu_{s}}^{\mu_{1}\cdots\mu_{r}} & = & -(\hat{X}_{\xi}\hat{T}_{\nu_{1}\cdots\nu_{n}}^{\mu_{1}\cdots\mu_{n}})+(\hat{X}_{(\partial_{\mu}\xi^{\mu_{1}})}\hat{T}_{\nu_{1}\cdots\nu_{n}}^{\mu\cdots\mu_{n}})+\cdots+(\hat{X}_{(\partial_{\mu}\xi^{\mu_{n}})}\hat{T}_{\nu_{1}\cdots\nu_{n}}^{\mu_{1}\cdots\mu})\\
 &  & -(\hat{X}_{(\partial_{\nu_{1}}\xi^{\nu})}\hat{T}_{\nu\cdots\nu_{n}}^{\mu_{1}\cdots\mu_{n}})-\cdots-(\hat{X}_{(\partial_{\nu_{n}}\xi^{\nu})}\hat{T}_{\nu_{1}\cdots\nu}^{\mu_{1}\cdots\mu_{n}}).\end{eqnarray*}

Up to now we have seen the following:

\begin{itemize}
\item Diffeomorphisms are generated by vector-fields $\xi\in\Xi$ and the
universal enveloping algebra ${\cal U}(\Xi)$ of the Lie algebra $\Xi$ of
vector-fields possesses a natural Hopf algebra structure defined by
(\ref{eq: HA of diffeos}). 
\item The algebra of scalar fields $\phi\in\mathcal{A}$ is a ${\cal U}(\Xi)$-module
algebra.
\item The universal enveloping algebra ${\cal U}(\Xi)$ can be deformed to a Hopf
algebra ${\cal U}(\hat{\Xi})$ defined in (\ref{eq: Deform diffeos alg},\ref{eq: deform diffeos coprod},\ref{eq: deform diffeos antipode and counit}). 
\item ${\cal U}(\hat{\Xi})$ consistently acts on the algebra of noncommutative functions
$\hat{\mathcal{A}}$, i.e. the algebra of noncommutative functions
is a ${\cal U}(\hat{\Xi})$-module algebra.
\item Regarding ${\cal U}(\hat{\Xi})$ as the underlying {}``symmetry'' of the
gravity theory to be built on the noncommutative space $\hat{\mathcal{A}}$,
we established a full tensor calculus as representations of the Hopf
algebra ${\cal U}(\hat{\Xi})$.
\end{itemize}

\section{Noncommutative Geometry}

The deformed algebra of infinitesimal diffeomorphisms and the tensor
calculus covariant with respect to it is the fundamental building-block
for the definition of a noncommutative geometry on $\theta-$deformed
spaces. In this section we sketch the important steps towards a deformed
Einstein-Hilbert action \cite{Aschieri:2005yw}. A first ingredient
is the \emph{covariant derivative} $\hat{D}_{\mu}$. Algebraically, it can
be defined by demanding that acting on a vector-field it produces
a tensor-field\begin{equation}
\hat{\delta}_{\xi}\hat{D}_{\mu}\hat{V}_{\nu}\stackrel{!}{=}-(\hat{X}_{\xi}\hat{D}_{\mu}\hat{V}_{\nu})-(\hat{X}_{(\partial_{\mu}\xi^{\alpha})}\hat{D}_{\alpha}\hat{V}_{\nu})-(\hat{X}_{(\partial_{\nu}\xi^{\alpha})}\hat{D}_{\mu}\hat{V}_{\alpha})\label{eq: cov deriv on vectro}\end{equation}
 The covariant derivative is given by a \emph{connection} $\hat{\Gamma}_{\mu\nu}{}^{\rho}$\[
\hat{D}_{\mu}\hat{V}_{\nu}=\hat{\partial}_{\mu}\hat{V}_{\nu}-\hat{\Gamma}_{\mu\nu}{}^{\rho}\hat{V}_{\rho}.\]
 From (\ref{eq: cov deriv on vectro}) it is possible to deduce the
transformation property of $\hat{\Gamma}_{\mu\nu}{}^{\rho}$\[
\hat{\delta}_{\xi}\hat{\Gamma}_{\mu\nu}{}^{\rho}=(\hat{X}_{\xi}\hat{\Gamma}_{\mu\nu}{}^{\rho})-(\hat{X}_{(\partial_{\mu}\xi^{\alpha})}\hat{\Gamma}_{\alpha\nu}{}^{\rho})-(\hat{X}_{(\partial_{\nu}\xi^{\alpha})}\hat{\Gamma}_{\mu\alpha}{}^{\rho})+(\hat{X}_{(\partial_{\alpha}\xi^{\rho})}\hat{\Gamma}_{\mu\nu}{}^{\alpha})-(\hat{\partial}_{\mu}\hat{\partial}_{\nu}\hat{\xi}^{\rho}).\]
The \emph{metric} $\hat{G}_{\mu\nu}$ is defined as a symmetric tensor
of rank two. It can be obtained for example by a set of \emph{}vector-fields
$\hat{E}_{\mu}{}^{a},\, a=0,\dots,3$\emph{,} where $a$ is to be
understood as a mere label. These vector-fields are called \emph{vierbeins.}
Then the symmetrized product of those vector-fields is indeed a symmetric
tensor of rank two\[
\hat{G}_{\mu\nu}:=\frac{1}{2}(\hat{E}_{\mu}{}^{a}\hat{E}_{\nu}{}^{b}+\hat{E}_{\nu}{}^{b}\hat{E}_{\mu}{}^{a})\eta_{ab}.\]
Here $\eta_{ab}$ stands for the usual flat Minkowski space metric
with signature $(-+++)$. Let us assume that we can choose the vierbeins
$\hat{E}_{\mu}{}^{a}$ such that they reduce in the commutative limit
to the usual vierbeins $e_{\mu}{}^{a}$. Then also the metric $\hat{G}_{\mu\nu}$
reduces to the usual, undeformed metric $g_{\mu\nu}$. 

The inverse metric tensor we denote by upper indices \[
\hat{G}_{\mu\nu}\hat{G}^{\nu\rho}=\delta_{\mu}^{\rho}.\]
 We use $\hat{G}_{\mu\nu}$ respectively $\hat{G}^{\mu\nu}$ to raise
and lower indices. 

The curvature and torsion tensors are obtained by taking the commutator
of two covariant derivatives%
\footnote{The generalization of covariant derivatives acting on tensors is straight
forward \cite{Aschieri:2005yw}.%
}

\[
[\hat{D}_{\mu},\hat{D}_{\nu}]\hat{V}_{\rho}=\hat{R}_{\mu\nu\rho}{}^{\alpha}\hat{V}_{\alpha}+\hat{T}_{\mu\nu}{}^{\alpha}\hat{D}_{\alpha}\hat{V}_{\rho}\]
 which leads to the expressions \begin{eqnarray*}
\hat{R}_{\mu\nu\rho}{}^{\sigma} & = & \hat{\partial}_{\nu}\hat{\Gamma}_{\mu\rho}{}^{\sigma}-\hat{\partial}_{\mu}\hat{\Gamma}_{\nu\rho}{}^{\sigma}+\hat{\Gamma}_{\nu\rho}{}^{\beta}\hat{\Gamma}_{\mu\beta}{}^{\sigma}-\hat{\Gamma}_{\mu\rho}{}^{\beta}\hat{\Gamma}_{\nu\beta}{}^{\sigma}\\
\hat{T}_{\mu\nu}{}^{\alpha} & = & \hat{\Gamma}_{\nu\mu}{}^{\alpha}-\hat{\Gamma}_{\nu\mu}{}^{\alpha}.\end{eqnarray*}
 If we assume the \emph{torsion-free} case, i.e. \[
\hat{\Gamma}_{\mu\nu}{}^{\sigma}=\hat{\Gamma}_{\nu\mu}{}^{\sigma},\]
 we find an unique expression for the metric connection (Christoffel
symbol) defined by \[
\hat{D}_{\alpha}\hat{G}_{\beta\gamma}\stackrel{!}{=}0\]
 in terms of the metric and its inverse%
\footnote{We don't introduce a new symbol for the metric connection.%
} \[
\hat{\Gamma}_{\alpha\beta}{}^{\sigma}=\frac{1}{2}(\hat{\partial}_{\alpha}\hat{G}_{\beta\gamma}+\hat{\partial}_{\beta}\hat{G}_{\alpha\gamma}-\hat{\partial}_{\gamma}\hat{G}_{\alpha\beta})\hat{G}^{\gamma\sigma}.\]

From the curvature tensor $\hat{R}_{\mu\nu\rho}{}^{\sigma}$ we get
the curvature scalar by contracting the indices\[
\hat{R}:=\hat{G}^{\mu\nu}\hat{R}_{\nu\mu\rho}{}^{\rho}.\]
 $\hat{R}$ indeed transforms as a scalar which may be checked explicitly
by taking the deformed coproduct (\ref{eq: deform diffeos coprod})
into account. 

To obtain an integral which is invariant with respect to the Hopf
algebra of deformed infinitesimal diffeomorphisms we need a measure
function $\hat{E}$. We demand the transformation property \begin{equation}
\hat{\delta}_{\xi}\hat{E}=-\hat{X}_{\xi}\hat{E}-\hat{X}_{(\partial_{\mu}\xi^{\mu})}\hat{E}.\label{eq: trafo of measur}\end{equation}
 Then it follows with the deformed coproduct (\ref{eq: deform diffeos coprod})
that for any scalar field $\hat{S}$\[
\hat{\delta}_{\xi}\hat{E}\hat{S}=-\hat{\partial}_{\mu}(\hat{X}_{\xi^{\mu}}(\hat{E}\hat{S})).\]
 Hence, transforming the product of an arbitrary scalar field with
a measure function $\hat{E}$ we obtain a total derivative which vanishes
under the integral. A suitable measure function with the desired transformation
property (\ref{eq: trafo of measur}) is for instance given by the
determinant of the vierbein $\hat{E}_{\mu}{}^{a}$ \[
\hat{E}=\mathrm{det}(\hat{E}_{\mu}{}^{a}):=\frac{1}{4!}\varepsilon^{\mu_{1}\cdots\mu_{4}}\varepsilon_{a_{1}\cdots a_{4}}\hat{E}_{\mu_{1}}{}^{a_{1}}\hat{E}_{\mu_{2}}{}^{a_{2}}\hat{E}_{\mu_{3}}{}^{a_{3}}\hat{E}_{\mu_{4}}{}^{a_{4}}.\]
 That $\hat{E}$ transforms correctly can be shown by using that the
product of four $\hat{E}_{\mu_{i}}{}^{a_{i}}$ transforms as a tensor
of fourth rank and some combinatorics.

Now we have all ingredients to write down an Einstein-Hilbert action.
Note that having chosen a differential calculus as in (\ref{def: differential calculus}),
the integral is uniquely determined up to a normalization factor by
requiring%
\footnote{We consider functions that {}``vanish at infinity''.%
}\cite{Douglas:2001ba} \[
\int\hat{\partial}_{\mu}\hat{f}=0\]
for all $\hat{f}\in\hat{\mathcal{A}}$. Then we define the \emph{Einstein-Hilbert
action} on $\hat{\mathcal{A}}$ as \[
\hat{S}_{\mathrm{EH}}:=\int\mathrm{det}(\hat{E}_{\mu}{}^{a})\hat{R}+\mathrm{complex\, conj.}.\]
 It is by construction invariant with respect to deformed diffeomorphisms
meaning that \[
\hat{\delta}_{\xi}\hat{S}_{\mathrm{EH}}=0.\]
 In this section we have presented the fundamentals of a noncommutative
geometry on the algebra $\hat{\mathcal{A}}$ and defined an invariant
Einstein-Hilbert action. There is however one important step missing
which is subject of the following section: We want to make contact
of the noncommutative gravity theory with Einstein's gravity theory.
This we achieve by introducing the $\star$-product formalism.

\section{Star Products and Expanded Einstein-Hilbert Action}

To express the noncommutative fields in terms of their commutative
counterparts we first observe that we can map the whole algebraic
construction of the previous sections to the algebra of commutative
functions via the vector space isomorphism $W$ introduced in Section
\ref{sec:Noncommutative-Spaces}. By equipping the algebra of commutative
functions with a new product denoted by $\star$ be can render $W$
an algebra isomorphism. We define \begin{equation}
f\star g:=W^{-1}(W(f)W(g))=W^{-1}(\hat{f}\hat{g})\label{def: star product}\end{equation}
 and obtain \[
(\mathcal{A},\star)\cong\hat{\mathcal{A}}.\]
 The $\star$-product corresponding to the symmetric ordering prescription
$W$ is then given explicitly by the Moyal-Weyl product%
\footnote{This is an immediate consequence of (\ref{eq: almost star product}).%
} \[
f\star g=\mu\circ e^{\frac{i}{2}\theta^{\mu\nu}\partial_{\mu}\otimes\partial_{\nu}}f\otimes g=fg+\frac{i}{2}\theta^{\mu\nu}(\partial_{\mu}f)(\partial_{\nu}g)+\mathcal{O}(\theta^{2}).\]
 It is a deformation of the commutative point-wise product to which
it reduces in the limit $\theta\rightarrow0$. 

In virtue of the isomorphism $W$ we can map all noncommutative fields
to commutative functions in $\mathcal{A}$\[
\hat{F}\mapsto W^{-1}(\hat{F})\equiv F.\]
 We then expand the image $F$ in orders of the deformation parameter
$\theta$\[
F=F^{(0)}+F^{(1)}+F^{(2)}+\mathcal{O}(\theta^{3}),\]
 where the zeroth order always corresponds to the undeformed quantity.
Products of functions in $\hat{\mathcal{A}}$ are simply mapped to
$\star$-products of the corresponding functions in $\mathcal{A}$.
The same can be done for the action of the derivative $\hat{\partial}_{\mu}$
and consequently for an arbitrary differential operator acting on
$\hat{\mathcal{A}}$ \cite{Aschieri:2005yw}. 

The fundamental dynamical field of our gravity theory is the vierbein
field $\hat{E}_{\mu}{}^{a}$. All other quantities such as metric,
connection and curvature can be expressed in terms of it. Its image
with respect to $W^{-1}$ is denoted by $E_{\mu}{}^{a}$. In first
approximation we study the case \[
E_{\mu}{}^{a}=e_{\mu}{}^{a},\]
 where $e_{\mu}{}^{a}$ is the usual vierbein field. Then for instance
the metric is given up to second order in $\theta$ by \begin{eqnarray*}
G_{\mu\nu} & = & \frac{1}{2}(E_{\mu}{}^{a}\star E_{\nu}{}^{b}+E_{\nu}{}^{b}\star E_{\mu}{}^{a})\eta_{ab}=\frac{1}{2}(e_{\mu}{}^{a}\star e_{\nu}{}^{b}+e_{\nu}{}^{b}\star e_{\mu}{}^{a})\eta_{ab}\\
 & = & g_{\mu\nu}-\frac{1}{8}\theta^{\alpha_{1}\beta_{1}}\theta^{\alpha_{2}\beta_{2}}(\partial_{\alpha_{1}}\partial_{\alpha_{2}}e_{\mu}{}^{a})(\partial_{\beta_{1}}\partial_{\beta_{2}}e_{\nu}{}^{b})\eta_{ab}+\dots,\end{eqnarray*}
where $g_{\mu\nu}$ is the usual, undeformed metric. For the Christoffel
symbol one finds up to second order: The zeroths order is the undeformed expression \begin{eqnarray} \Gamma_{\mu\nu}^{(0)\rho} \hspace*{-2mm}&=&\hspace*{-2mm}  \frac{1}{2} \big( \p_\mu g_{\nu\gamma} + \p_\nu g_{\mu\gamma}  - \p_\gamma g_{\mu\nu}\big) g^{\gamma\rho} ,\label{10.5} \end{eqnarray} the first order reads \begin{equation} \Gamma^{(1)}_{\mu\nu}{}^{\rho} = \frac{i}{2}\theta^{\alpha\beta}(\partial_{\alpha}\Gamma_{\mu\nu}^{(0)\sigma})g_{\sigma\tau} (\partial_{\beta}g^{\tau\rho}) \end{equation} and the second order \begin{eqnarray} \Gamma^{(2)}_{\mu\nu}{}^{\rho} =   &=&\hspace*{-2mm} - \frac{1}{8}\theta^{\alpha_{1}\beta_{1}}\theta^{\alpha_{2}\beta_{2}}\bigg( (\partial_{\alpha_{1}}\partial_{\alpha_{2}}\Gamma_{\mu\nu\sigma}^{(0)}) (\partial_{\beta_{1}}\partial_{\beta_{2}}g^{\sigma\rho}) -2(\partial_{\alpha_{1}}\Gamma_{\mu\nu\sigma}^{(0)})\partial_{\beta_{1}}\big((\partial_{\alpha_{2}}g^{\sigma\tau})(\partial_{\beta_{2}}g_{\tau\xi})g^{\xi\rho}\big) \nonumber\\ &-&\hspace*{-2mm}\Gamma_{\mu\nu\sigma}^{(0)}\Big( (\partial_{\alpha_{1}}\partial_{\alpha_{2}}g^{\sigma\tau})(\partial_{\beta_{1}}\partial_{\beta_{2}}g_{\tau\xi}) +g^{\sigma\tau}(\partial_{\alpha_{1}}\partial_{\alpha_{2}}e_{\tau}{}^{a}) (\partial_{\beta_{1}}\partial_{\beta_{2}}e_{\xi}{}^{b})\eta_{ab} \nonumber\\ &-&\hspace*{-2mm}2\partial_{\alpha_{1}}\big( (\partial_{\alpha_{2}}g^{\sigma\tau})(\partial_{\beta_{2}}g_{\tau\lambda})g^{\lambda\kappa}\big) (\partial_{\beta_{1}}g_{\kappa\xi}) \Big) g^{\xi\rho}  +\frac{1}{2}\Big( \partial_{\mu}\big( (\partial_{\alpha_{1}}\partial_{\alpha_{2}}e_{\nu}^{\ a}) (\partial_{\beta_{1}}\partial_{\beta_{2}}e_{\sigma}^{\ b})\big) \nonumber\\ &+&\hspace*{-2mm}\partial_{\nu}\big( (\partial_{\alpha_{1}}\partial_{\alpha_{2}}e_{\sigma}^{\ a}) (\partial_{\beta_{1}}\partial_{\beta_{2}}e_{\mu}^{\ b})\big) -\partial_{\sigma}\big( (\partial_{\alpha_{1}}\partial_{\alpha_{2}}e_{\mu}^{\ a}) (\partial_{\beta_{1}}\partial_{\beta_{2}}e_{\nu}^{\ b}) \big) \Big) \eta_{ab}g^{\sigma\rho}\bigg), \label{10.6} \end{eqnarray} where \begin{equation} \Gamma_{\mu\nu\sigma}^{(0)} = \Gamma_{\mu\nu}^{(0)\rho}g_{\rho\sigma} . \end{equation}

The expressions for the curvature tensor read \begin{eqnarray} R_{\mu\nu\rho}^{(1)}{}^{\sigma} \hspace*{-2mm}&=&\hspace*{-2mm} -\frac{i}{2}\theta^{\kappa\lambda}\bigg( (\partial_{\kappa}R_{\mu\nu\rho}^{(0)}{}^{\tau}) (\partial_{\lambda}g_{\tau\gamma})g^{\gamma\sigma}  -(\partial_{\kappa}\Gamma_{\nu\rho}^{(0)}{}^{\beta})\Big( \Gamma_{\mu\beta}^{(0)}{}^{\tau}(\partial_{\lambda}g_{\tau\gamma})g^{\gamma\sigma} \nonumber\\ &&-\Gamma_{\mu\tau}^{(0)}{}^{\sigma}(\partial_{\lambda}g_{\beta\gamma})g^{\gamma\tau} +\partial_{\mu}\big( (\partial_{\lambda}g_{\beta\gamma})g^{\gamma\sigma}\big) +(\partial_{\lambda}\Gamma_{\mu\beta}^{(0)}{}^{\sigma})\Big) \nonumber\\ && +(\partial_{\kappa}\Gamma_{\mu\rho}^{(0)}{}^{\beta})\Big( \Gamma_{\nu\beta}^{(0)}{}^{\tau}(\partial_{\lambda}g_{\tau\gamma})g^{\gamma\sigma} -\Gamma_{\nu\tau}^{(0)}{}^{\sigma}(\partial_{\lambda}g_{\beta\gamma})g^{\gamma\tau} \nonumber\\ &&+\partial_{\nu}\big( (\partial_{\lambda}g_{\beta\gamma})g^{\gamma\sigma}\big) +(\partial_{\lambda}\Gamma_{\nu\beta}^{(0)}{}^{\sigma})\Big) \bigg) \label{10.7}\\ R_{\mu\nu\rho}^{(2)}{}^{\sigma} \hspace*{-2mm}&=&\hspace*{-2mm} \partial_{\nu}\Gamma_{\mu\rho}^{(2)}{}^{\sigma} +\Gamma_{\nu\rho}^{(2)}{}^{\gamma}\Gamma_{\mu\gamma}^{(0)}{}^{\sigma} +\Gamma_{\nu\rho}^{(0)}{}^{\gamma}\Gamma_{\mu\gamma}^{(2)}{}^{\sigma} \nonumber\\ && + \frac{i}{2}\theta^{\alpha\beta}\Big( (\partial_{\alpha}\Gamma_{\nu\rho}^{(1)}{}^{\gamma})(\partial_{\beta}\Gamma_{\mu\gamma}^{(0)}{}^{\sigma}) +(\partial_{\alpha}\Gamma_{\nu\rho}^{(0)}{}^{\gamma})(\partial_{\beta}\Gamma_{\mu\gamma}^{(1)}{}^{\sigma}) \Bigr) \nonumber\\ && - \frac{1}{8}\theta^{\alpha_{1}\beta_{1}}\theta^{\alpha_{2}\beta_{2}} (\partial_{\alpha_{1}}\partial_{\alpha_{2}} \Gamma_{\nu\rho}^{(0)}{}^{\gamma}) (\partial_{\beta_{1}}\partial_{\beta_{2}}\Gamma_{\mu\gamma}^{(0)}{}^{\sigma}) \quad -(\mu\leftrightarrow\nu) ,\label{10.8} \end{eqnarray}where
the second order is given implicitly in terms of the Christoffel symbol. 

The deformed Einstein-Hilbert action is given by \begin{eqnarray}
S_{\mathrm{EH}} & = & \frac{1}{2}\int\mathrm{d}^{4}x\,\mathrm{det}_{\star}e_{\mu}{}^{a}\star R+\mathrm{c.c.}\nonumber \\
 & = & \frac{1}{2}\int\mathrm{d}^{4}x\,\mathrm{det}_{\star}e_{\mu}{}^{a}\star(R+\bar{R})\nonumber \\
 & = & \frac{1}{2}\int\mathrm{d}^{4}x\,\mathrm{det}_{\star}e_{\mu}{}^{a}(R+\bar{R})\nonumber \\
 & = & S_{\mathrm{EH}}^{(0)}+\int\mathrm{d}^{4}x\,(\mathrm{det}e_{\mu}{}^{a})R^{(2)}+(\mathrm{det}_{\star}e_{\mu}{}^{a})^{(2)}R^{(0)},\label{eq: deformed Einstein-Hilbert action}\end{eqnarray}
 where we used that the integral together with the Moyal-Weyl product
has the property%
\footnote{This follows by partial integration.%
}\[
\int\mathrm{d}^{4}x\, f\star g=\int\mathrm{d}^{4}x\, fg=\int\mathrm{d}^{4}x\, g\star f.\]
 In (\ref{eq: deformed Einstein-Hilbert action}) $\mathrm{det}_{\star}e_{\mu}{}^{a}$
is the $\star$-determinant \begin{eqnarray*}
\mathrm{det}_{\star}e_{\mu}{}^{a} & = & \frac{1}{4!}\varepsilon^{\mu_{1}\cdots\mu_{4}}\varepsilon_{a_{1}\cdots a_{4}}e_{\mu_{1}}{}^{a_{1}}\star e_{\mu_{2}}{}^{a_{2}}\star e_{\mu_{3}}{}^{a_{3}}\star e_{\mu_{4}}{}^{a_{4}}\\
 & = & \mathrm{det}e_{\mu}{}^{a}+(\mathrm{det}_{\star}e_{\mu}{}^{a})^{(2)}+\dots,\end{eqnarray*}
where \begin{eqnarray} (\mathrm{det}_\star)^{(2)} &=& - \frac{1}{8}\frac{1}{4!}\theta^{\alpha_{1}\beta_{1}}\theta^{\alpha_{2}\beta_{2}} \varepsilon^{\mu_{1}\dots\mu_{4}}\varepsilon_{a_{1}\dots a_{4}} \nonumber\\ &&\Big( (\partial_{\alpha_{1}}\partial_{\alpha_{2}}e_{\mu_{1}}{}^{a_{1}}) (\partial_{\beta_{1}}\partial_{\beta_{2}}e_{\mu_{2}}{}^{a_{2}})e_{\mu_{3}}{}^{a_{3}}e_{\mu_{4}}{}^{a_{4}} \nonumber\\ &&+ \,\partial_{\alpha_{1}}\partial_{\alpha_{2}}(e_{\mu_{1}}{}^{a_{1}}e_{\mu_{2}}{}^{a_{2}}) (\partial_{\beta_{1}}\partial_{\beta_{2}}e_{\mu_{3}}{}^{a_{3}})e_{\mu_{4}}{}^{a_{4}} \nonumber\\ &&+\,\partial_{\alpha_{1}}\partial_{\alpha_{2}}(e_{\mu_{1}}{}^{a_{1}}e_{\mu_{2}}{}^{a_{2}}e_{\mu_{3}}{}^{a_{3}}) (\partial_{\beta_{1}}\partial_{\beta_{2}}e_{\mu_{4}}{}^{a_{4}})\Big) .\label{10.12} \end{eqnarray}The
odd orders of $\theta$ vanish in (\ref{eq: deformed Einstein-Hilbert action})
but the even orders of $\theta$ give nontrivial contributions.

Equation (\ref{eq: deformed Einstein-Hilbert action}) shows explicitly
the corrections to Einsteins gravity predicted by the noncommutative
theory.

\subsection*{Remarks}

For an introduction to field theories on noncommutative spaces, we
recommend the review articles \cite{Douglas:2001ba,Szabo:2001kg}.
To learn more about related approaches to noncommutative geometry
the reader is referred to \cite{Connes,Madore:2000aq}. More about
Hopf algebras and Quantum Groups can be found in \cite{Chari-Presley,Klimyk:1997eb,Majid}.
A good pedagogical introduction to $\star$-products can be found
in \cite{Waldmann}. The construction of a gravity theory presented in this lecture is based on
\cite{Aschieri:2005yw,Aschieri:2005zs}. 



\section*{Acknowledgments}

The results presented in this lecture were
obtained together with P. Aschieri, C. Blohmann, M. Dimitrijevi{\' c},
P. Schupp and J. Wess and I would like to thank them for their collaboration
from which I learned so much.

\bibliographystyle{diss}
\bibliography{literature_gravity}

\end{document}